\documentclass[9pt,twocolumn,twoside]{osajnl}

\usepackage{graphicx}
\usepackage{booktabs,multirow}
\usepackage{amsbsy,amsmath}
\usepackage{amssymb}
\usepackage{mathtools}
\journal{ol} 

\setboolean{shortarticle}{true} 

\ifthenelse{\boolean{shortarticle}}{\colorlet{color2}{color2b}}{\colorlet{color2}{color2}}

\title{Polynomials of Gaussians and vortex-Gaussian beams as complete, transversely confined bases
}

\author[*]{Rodrigo Guti\'{e}rrez-Cuevas}
\author[ ]{Miguel A. Alonso}

\affil[ ]{Center for Coherence and Quantum Optics and The Institute of Optics, University of Rochester, Rochester, New York 14627, USA}

\affil[*]{Corresponding author: rgutier2@ur.rochester.edu}

\dates{Compiled \today}

\ociscodes{(260.1960 )  Diffraction theory; (260.0260)   Physical optics; (260.6042) Singular optics; (000.3860)  Mathematical methods in physics.}

\doi{\url{http://dx.doi.org/10.1364/ao.XX.XXXXXX}}

\begin{abstract}
A novel type of discrete basis for paraxial beams is proposed, consisting of monomial vortices times polynomials of Gaussians in the radial variable. These bases have the distinctive property that the effective size of their elements is roughly independent of element order, meaning that the optimal scaling for expanding a localized field does not depend significantly on truncation order. This behavior contrasts with that of bases composed of polynomials times Gaussians, such as Hermite-Gauss and Laguerre-Gauss modes, 
where the scaling changes roughly as the inverse square root of the truncation order. 
\end{abstract}

\setboolean{displaycopyright}{true}

\begin{document}
\newcommand{\pd}[2]{\frac{\partial #1}{\partial #2}} 
\newcommand{\ket}[1]{\left| #1 \right>} 
\newcommand{\bra}[1]{\left< #1 \right|} 
\newcommand{\bs}{\boldsymbol}
\newcommand{\sech}{\text{sech}}

\maketitle

\thispagestyle{fancy}
\ifthenelse{\boolean{shortarticle}}{\ifthenelse{\boolean{singlecolumn}}{\abscontentformatted}{\abscontent}}{}


Most light beams propagating through free space are not described by closed-form expressions. Instead, they are modeled through numerical evaluation of the angular spectrum or Fresnel integrals, which correspond to expansions in terms of continuous basis sets whose elements are plane and (paraxial approximations to) spherical waves, respectively. In some cases, the numerical computation can be simplified if a discrete basis expansion of the field is used instead, as long as the elements have simple closed-form expressions. In general, the geometry of the problem dictates which basis to use. For example, if the problem lends itself to Cartesian coordinates, then Hermite-Gauss (HG) beams \cite{siegman86lasers,alonso2012basis} are a natural choice. On the other hand, when studying beams with rotational symmetry, including those carrying orbital angular momentum \cite{yao2011orbital,andrews2012angular}, then  Laguerre-Gauss (LG) beams \cite{siegman86lasers,alonso2012basis} are more suitable. More exotic geometries possess their own type of natural separable basis (see for example \cite{bandres2004bince}).
Within a given geometry, however, there are other possible bases to choose from, e.~g.~ the elegant modifications to the bases just mentioned \cite{siegman86lasers,takenaka1985propagation,bandres2004elegant}, which give up  orthogonality in favor of ease in notation. 
Note that orthogonality is usually desirable but not required; completeness, on the other hand, is needed. 
Other types of beams have been proposed, such as the polynomial Gaussian beams \cite{roux2006polynomial,roux2008topological} and the scattering modes \cite{ferrando2016analytical} which allow the study and propagation of beams with more complicated phase singularities at the initial plane.

All the bases just mentioned include a Gaussian factor to limit the spatial extent of their elements to a region where the field is localized. A Gaussian is convenient because Gaussian beams are the simplest finite-power (square-integrable) solution to the paraxial wave equation. To form the basis, this Gaussian is multiplied by polynomials in the coordinates, which can be made to be orthogonal when using a weight function involving the Gaussian. These polynomial factors widen the elements roughly proportionally to the square root of their order. 

Here, we propose using a different approach: rather than using it as a weight factor, we build new bases by letting the Gaussian be the argument of the polynomials. That is, instead of polynomials \emph{times} a Gaussian, the new bases use polynomials \emph{of} Gaussians, the polynomial variable being $u=\exp(-r^2)\in[0,1]$, 
where $r$ is the (scaled) radial variable in polar coordinates at the initial plane. 
Like for LG beams, the order $n$ of the polynomial indicates the number of radial zeros for the new basis elements, and 
the azimuthal structure is achieved by a simple vortex factor $r^{|m|}\exp(im\phi)$. Each basis element is then a combination of simple vortex-Gaussian beams that are self-Fourier objects and self-similar under paraxial propagation. 

Perhaps the main feature of the new bases is that all their elements have essentially equal effective width, which stands in sharp contrast with standard bases like LG modes for which the width of the elements increases with order. That is, while the LG modes are the natural analogs of the two-dimensional harmonic oscillator, the new basis elements have more in common qualitatively with the eigenfunctions of an infinite well potential or the modes of a drum skin, which are all constrained to the same region irrespective of mode order. As a consequence of this property, the optimal transverse scaling for fitting a well-localized beam with a finite number of elements is roughly independent of truncation order.

We propose three bases of this type. Two of them are based on standard polynomials so they are easiest to implement. The first, however, has elements that are not easy to propagate, so it is discarded. The second resolves this issue at the cost of its orthogonality relation involving extra weight functions. The third is strictly orthonormal with uniform weight, but requires the definition of new orthogonal polynomials.
These bases are, in turn, compared with the standard LG basis through a couple of examples that illustrate their main properties. 
%


\emph{Gauss-Legendre (GL) basis.}---In previous work, 
an orthonormal basis for nonparaxial fields with rotational symmetry
was introduced \cite{alonso2006new}. 
In the paraxial limit, the extension of this basis to general fields was found to reduce to the form
\begin{align}
\label{eq:lg}
B_{n,m}(r, \phi)= b_{n,m} e^{-\frac{r^2}{2}} \bar P_{|m|+n}^{(m)}(e^{-r^2})e^{im\phi},
\end{align}
where $\bar P_{|m|+n}^{(m)}(u)=P_{|m|+n}^{(m)}(2u-1)$ is the shifted associated Legendre function, $n=0,1,...$ is the number of radial nodes, $m$ is the vorticity (which can take any integer value irrespective of $n$), and the normalization factor is given by
\begin{align}
b_{n,m}=\sqrt{\frac{(2|m|+2n+1)(|m|+n-m)!}{\pi (|m|+n+m)!}}.
\end{align}
Here, we consider a normalized radial variable, $r=\rho/\alpha$, where $\rho$ is the radial coordinate and $\alpha$ is a width scaling parameter of the basis elements. By using the change of variables $u=\exp(-r^2)$, it is easy to verify that the functions in Eq.~(\ref{eq:lg}) are in fact orthonormal over the plane.

For $m=0$ these functions become sums of Gaussians with different widths, and therefore provide a formalization of the approach in \citep{wen1988diffraction,ding2004notes}, where an optimization procedure was used to find the best fit of Gaussians to any axis-symmetric beam. By instead using the orthonormal basis elements, the coefficients can be determined through simple inner products. For a general integer $m$, Eq.~(\ref{eq:lg}) presents an extension to beams without rotational symmetry. However, 
as discussed in \cite{alonso2006new}, the propagation of these basis elements away from the initial plane is not given by a simple closed-form expression because 
%
%
the associated Legendre functions are not polynomials; for odd $m$, these functions include a fractional power factor. 


\begin{figure}
\centering
\includegraphics[width=.450\linewidth]{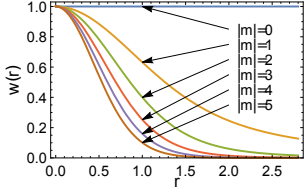}
\caption{\label{fig:wparax} Weight function, Eq.~(\ref{eq:wparax}), for different values of $m$. 
}
\end{figure}

\emph{Modified Gauss-Legendre (MGL).}---
We now define a second basis by eliminating the problematic fractional powers,
with the compromise of modifying the orthogonality relation over the plane. The basis elements are
\begin{align}
\label{eq:mlg}
 M_{n,m}(r, \phi)= b_{n,m}  e^{-\frac{r^2}{2}}\left[ \frac{\bar P_{|m|+n}^{(m)}\left(e^{-r^2}\right)}{\left(1-e^{-r^2}\right)^{|m|/2}} \right]r^{|m|}e^{im\phi},
\end{align}
where the product in square brackets is a polynomial of the Gaussian, which (if preferred) can be written in terms of the Jacobi polynomial as $\exp(-|m|r^2/2)P_{n}^{(m,m)}[2\exp(-r^2)-1]$. 
These elements are orthonormal with respect to the weight function
\begin{align}
\label{eq:wparax}
W_m(r)= \left[(1-e^{-r^2})/r^2 \right]^{|m|}.
\end{align}
Figure \ref{fig:wparax} depicts this radial weight function for different values of $m$. Notice that as $|m|$ increases, 
the region considered for orthogonality reduces to an increasingly localized patch near the origin. 
We can expand any function in terms of this basis as
\begin{align}
U(r,\phi)=\sum_{n=0}^\infty\sum_{m=-\infty}^{\infty} a_{n,m} M_{n,m}(r,\phi),
\end{align}
where the coefficients are given by
\begin{align}
a_{n,m}=\int_0^\infty\int_0^{2\pi}U(r,\phi)M^*_{n,m}(r,\phi)W_{m}(r)r d\phi dr.
\end{align}
Given the nonuniform weight function, this basis 
does not guarantee a decrease in rms error (uniformly weighted over the plane) when adding more terms. 
However, 
because the weight is localized where the fitted functions are expected to be most important, the accuracy of the fit is comparable to that for a similar orthonormal basis. Note that this basis coincides with the GL basis for $m=0$ and therefore it is orthogonal in this case.

\begin{figure}
\centering
\includegraphics[width=1\linewidth]{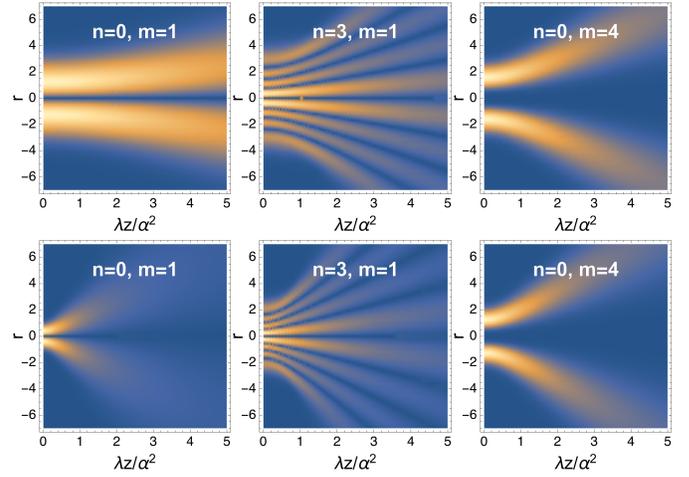}
\caption{\label{fig:modprop} Spatial propagation of the intensity of the MGL basis elements for different values of $m$ and $n$. The first (second) row shows the propagation for fields  whose initial position (Fourier) distribution is given by Eq.~(\ref{eq:mlg}). 
}
\end{figure}

Since each
basis element is a finite sum of Gaussians with different widths multiplied by simple vortices,
the Fourier transform (from $r,\phi$ to $f,\theta$) 
can be computed via the formula
\begin{align}
\label{eq:selfsim}
\mathcal F\{r^me^{-q r^2} e^{im\phi}\}
=& (-i)^m  \left(\pi/q\right)^{m+1} e^{im\theta}f^m e^{-\pi^2 f^2/q}, 
\end{align}
thus allowing the simple propagation of its elements in closed form and without the use of special functions.  Note that this basis can be used to express fields prescribed in position space as well as in Fourier space (angular spectrum). Figure \ref{fig:modprop} shows how some elements of the basis propagate when used in position and Fourier spaces, respectively.



\emph{Gauss-New (GN).}---An aspect of the MGL basis that is not ideal is the need of a nonuniform spatial weight function for orthogonality. We can, however, use it as a guideline to define an orthogonal basis with uniform spatial weight. 
This new basis has the same structure as that in Eq.~(\ref{eq:mlg}), 
but with new polynomials replacing the Jacobi polynomials:
\begin{align}
\mathcal D_{n,m}(r, \phi)= d_{n,m} e^{-\frac{r^2}{2}} e^{-\frac{|m|}{2}r^2} D_n^{(m)}\left(e^{-r^2}\right)r^{|m|} e^{im\phi},
\end{align}
where the polynomials $D_n^{(m)}$ are defined such that
\begin{align}
\int_0^\infty\int_0^{2\pi}\mathcal{D}_{n,m}(r,\phi)\mathcal{D}^*_{n',m'}(r,\phi)r d\phi dr= \delta_{n,n'}\delta_{m,m'}.
\end{align}
As the angular part immediately gives orthogonality for $m\neq m'$, we only need to worry about the radial part for $m=m'$. The change of variable $u=\exp(-r^2)$ leads to the following orthogonality condition that the $D_{n}^{(m)}$ polynomials must satisfy:
\begin{align}
 \int_0^1 D_n^{(m)}(u)D_{n'}^{(m)}{}^*(u)
 w_m(u)
 du= \frac{\delta_{n,n'}}{\pi |d_{n,m}|^2},
\end{align}
with the weight in the space of $u$ given by 
\begin{align}
w_m(u)=\left(-\ln u\right)^{|m|}u^{|m|}.
\end{align}
\begin{figure}
\centering
\includegraphics[width=.95\linewidth]{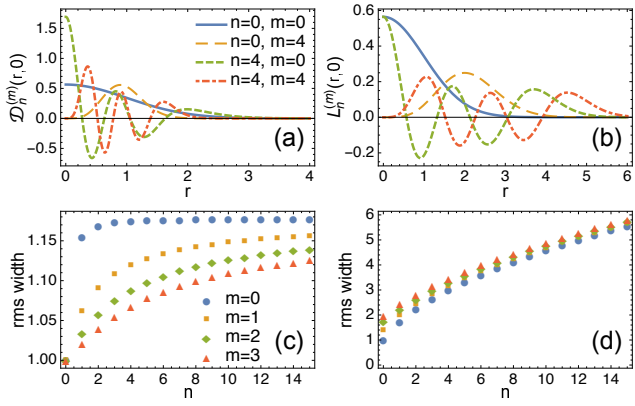}
\caption{\label{fig:rplots} Radial dependence of the (a) GN and (b) LG bases 
and their respective rms width (c) and (d) for different $m$ and $n$.}
\end{figure}
Luckily, there is a standard method \cite{szego1967orthogonal,alonso2012basis} to construct these polynomials 
in terms of their moments:
\begin{align}
\label{eq:detpol}
D_n^{(m)}(u)=\left| 
\begin{array}{cccc}
\mu_0^{(m)}&\mu_1^{(m)}& \cdots &\mu_n^{(m)}\\
\mu_1^{(m)}&\mu_2^{(m)}& \cdots &\mu_{n+1}^{(m)}\\
\vdots& \vdots & \ddots  & \vdots\\
\mu_{n-1}^{(m)}&\mu_n^{(m)}& \cdots &\mu_{2n-1}^{(m)}\\
1&u& \cdots &u^{n}
\end{array}
\right|
\end{align}
where the moments $\mu_n^{(m)}$ in this case have the simple form
\begin{align}
\label{eq:mom}
\mu_n^{(m)}=\int_0^1w_m(u)u^n du=\frac{|m|!}{(n+|m|+1)^{|m|+1}}.
\end{align}
Note that the factor of $|m|!$ can be taken out of the determinant in order to simplify the calculations. The normalization is given by $d_{n,m}=[\pi \,h_n^{(m)}]^{-1/2}$,
where $h_n^{(m)}$ is the norm of the $D_{n}^{(m)}$ polynomial which is given by $h_n^{(m)}= \Delta^{(m)}_{n-1} \Delta^{(m)}_n$ with
 \begin{align}
\Delta^{(m)}_n=\left|
\begin{array}{cccc}
\mu^{(m)}_0&\mu^{(m)}_1& \cdots &\mu^{(m)}_n\\
\mu^{(m)}_1&\mu^{(m)}_2& \cdots &\mu^{(m)}_{n+1}\\
\vdots& \vdots & \ddots  & \vdots\\
\mu^{(m)}_{n}&\mu^{(m)}_{n+1}& \cdots &\mu^{(m)}_{2n}
\end{array}
\right|.
\end{align}

As mentioned at the outset, the defining feature of the new bases is that their elements maintain the same rough width, independently of mode order. This feature is shown in Fig.~\ref{fig:rplots} where we plot the radial dependence and rms width for the GN functions, as well as for the LG modes for contrast. (The MGL are not shown as they are very similar to the GN functions.) 
\begin{figure}
\centering
\includegraphics[width=.95\linewidth]{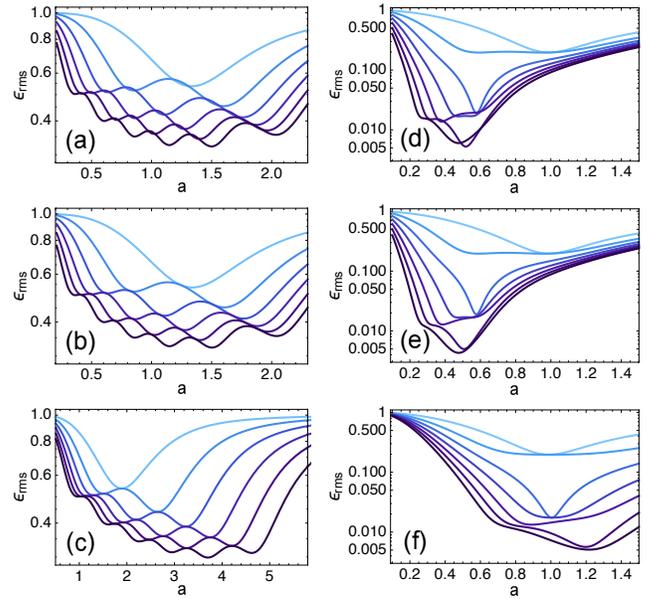}
\caption{\label{fig:errVor} (First column) Rms error for the expansion of a simple vortex of charge $m=1$ delimited by a circular aperture, as a function of its width when using (a) the MGL basis, (b) the GN basis, and (c) the LG basis, the number of terms used in the approximation ranges from 1 (lighter curves) to 8 (darker curves). (Second column) Rms error for the Gaussian field with $\phi$-dependent width as defined in Eq.~(\ref{eq:ell}) with $\delta =0.2$, as a function of the width parameter $a$ for (d) the MLG, (e) the GN basis, and (f) the LG for truncation orders $l_{\text{max}}=0$ (lighter curves) to $l_{\text{max}}=6$ (darker curves), where $l=|m|+n$.}
\end{figure}
Like the MGL basis, the elements of this basis have a simple Fourier transform and can be propagated paraxially in closed form, but they offer the advantages of an orthonormal basis. We do not present plots for their propagation as they look very similar to those in Fig.~\ref{fig:modprop}. The only slight disadvantage of this basis is the need to construct the new set of polynomials $D_{n}^{(m)}$. 


\emph{Comparison: Fitting prescribed fields.}---
We now compare the performance of the MGL and GN bases when approximating different fields with a finite number of elements. We also include the corresponding results for the standard LG basis. As the new bases coincide for $m=0$, we only consider non-rotationally symmetric fields.

\begin{figure}
\centering
\includegraphics[width=.95\linewidth]{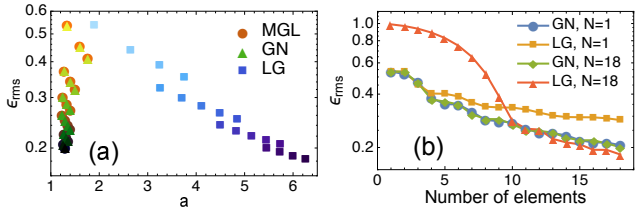}
\caption{\label{fig:minerr} (a) Minimum truncation error as a function of scaling parameter $a$ for the MGL, GN and LG bases for orders raging from 1 (lighter points) to 18 (darker points). (b) Truncation error as a function of elements in the approximation given a scaling parameter optimized for $N$ elements.}
\end{figure}

We start by considering simple vortices delimited by a circular aperture given by
\begin{align}\label{eq:aperture}
U_m(r,\phi)=\left(r/a\right)^{|m| }e^{im\phi} \text{circ} \left(r/a\right).
\end{align}
We use this example given its simplicity. However, because of the discontinuity at $r=a$ the convergence for all bases is slow, with the error being roughly inversely proportional to the truncation order.
In Fig.~\ref{fig:errVor}(a-c) we show the results for $m=1$. We see that the minimum error is localized within a small region for the MGL and GN bases, whereas for the LG basis it shifts towards higher values of the scaling parameter $a$ as the order increases (note the difference in ranges). Figure \ref{fig:minerr}(a) clarifies this behavior by considering larger truncation orders and showing only the minimum error and the corresponding value of $a$. It is clear that the value of $a$ that achieves the minimum error is roughly constant for the MGL and GN bases (and the values are very similar for both bases), with only slight shifts due to the small oscillations of the error curves [visible in Fig.~\ref{fig:errVor}(a-c)]. For the LG basis, on the other hand, the optimal $a$ shifts towards higher values. 

Notice that, for this first example, the truncation errors are slightly smaller for the LG basis than for the new bases. 
%
However, the advantage of having a well-localized optimal scaling can be appreciated when looking at the truncation error for fixed $a$.  That is, suppose that a value of $a$ is used that is optimal for some truncation order $N$, but then less or more terms are used. This is shown in Fig.~\ref{fig:minerr}(b) for the GN and LG bases, where the points labeled $N=1$ and $N=18$ show the rms error varying with truncation order, having chosen a scaling parameter that minimizes the error for either 1 and 18 terms, respectively. For the GN basis, the optimal $a$ in both cases is essentially the same, and so are the curves for the decay in error with truncation order. For the LG basis, on the other hand, if we choose $a$ to be optimal for a small truncation order, then the error reduces slowly when more terms are added, or if we use a scaling that optimizes convergence when using many elements (like $N=18$), the first few terms of the expansion on their own do a poor job at matching the desired function.

%
%
It is also worth mentioning that, while the nonorthogonal MGL basis does not guarantee a decrease in the error when adding terms for specific values of $a$ [as can be seen in Fig.~\ref{fig:errVor}(a) by the crossings in the curves], overall the error diminishes with the number of terms and attains minimum values comparable to those of its orthogonal counterparts [see Fig.~\ref{fig:minerr}(a) where the data points for the MGL and GN bases overlap almost perfectly]. 
For higher order vortices ($|m|\ge2$), the truncation error behaves in similar way as for $m=1$, the only noticeable difference being that for increasing $m$ the nonorthogonality of the MLG basis becomes more apparent. When plotting the truncation error as a function of $a$ the crossings are enhanced by an increase in the amplitude of the oscillation and the minimum value attained starts differing more from that of its orthogonal counterpart. 

As a second example, consider a Gaussian field with a $\phi$-dependent waist, 
\begin{align}
\label{eq:ell}
U(r,\phi)=\exp\left[{-\frac{r^2}{2a^2(1+\delta  \cos 2\phi)^2}}\right].
\end{align}
Due to its symmetry, only even $m$ terms are needed. The smoothness of this function means that the errors attained by all bases are significantly smaller.
The results are shown in Fig.~\ref{fig:errVor}(d-f), where we can again appreciate the rough invariance of the position of the optimal scaling with truncation order.  Here, we considered a total truncation order $l_{\rm max}$ for the index $l=n+|m|$ which is increased by unity for each curve. That is, for given $l_{\rm max}$, the number of elements used is $(l_{\rm max}+1)^2$. Note that, unlike in the previous example, for this case the GN basis attains a smaller truncation error than the LG basis.
\emph{Concluding remarks.}---We presented alternatives to the well-established LG basis consisting of polynomials of Gaussians with well defined vorticity. This approach allows basis elements with the same type of structure but with different, interesting properties: the simple form of the constituents (Gaussians and vortex Gaussians) makes it easy to propagate the elements, and these elements maintain the same effective size for all orders. This second property leads to a roughly order-independent scaling width for minimizing truncation error, allowing the reduction of the search-space for the optimal fit, and not requiring a new optimization if different truncation orders are used for different purposes. The non-orthogonal MGL basis provides accurate fits while allowing the use of standard polynomials that are inbuilt in common programming environments.  

Let us finish by discussing future generalizations of the work proposed here. First, analogous bases for one dimension could be obtained, where instead of elements with different vorticity, one considers even and odd functions. The resulting bases would constitute an alternative to the HG modes, and like these modes they can be used to write two-dimensional basis functions that are separable over the initial plane and follow simple propagation rules. A second generalization that turns out to be fairly straightforward is that into the nonparaxial regime, both for scalar and electromagnetic fields \cite{gutierrez-cuevas2017scalar}. We expect that this generalization will be particularly convenient for the study of the scattering of focused fields off spherical particles. Such extensions will be discussed in future work. 
\\

\noindent \textbf{Funding.} National Science Foundation (NSF) (PHY-1507278), CONACYT fellowship awarded to RGC.\\


\end{document}